# Control of Decoherence and Correlation in Single Quantum Dissipative Oscillator Systems

R. W. Rendell and A. K. Rajagopal
Naval Research Laboratory, Washington, DC 20375-5320

## ABSTRACT

A single quantum dissipative oscillator described by the Lindblad equation serves as a model for a nanosystem. This model is solved exactly by using the ambiguity function. The solution shows the features of decoherence (spatial extent of quantum behavior), correlation (spatial scale over which the system localizes to its physical dimensions), and mixing (mixed-state spatial correlation). A new relation between these length scales is obtained here. By varying the parameters contained in the Lindblad equation, it is shown that decoherence and correlation can be controlled. We indicate possible interpretation of the Lindblad parameters in the context of experiments using engineered reservoirs.

A prototypical model of many physical systems is often the quantum dissipative oscillator. This model contains the physical ideas of decoherence, correlation, diffusion, etc. associated with quantum systems. This is especially pertinent in discussing quantum nanodevice systems, which are usually imbedded in other systems so that a suitable description of the environmental effects may be described in terms of the parameters of such a model. The equation governing the density matrix of this system is the Lindblad equation, which describes dissipative quantum systems in a consistent way; namely, its solution does not violate basic requirements of positive semi-definiteness, probability conservation, and hermiticity of the density matrix [1]. The purpose of this paper is to exhibit this model in general terms of a Gaussian density matrix, which contains all the above elements. The "ambiguity function" [2] introduced originally in radar theory, defined as the Fourier transform in the center-of-mass coordinate of the density matrix is found to lead to solutions of the Lindblad's quantum dissipative oscillator. We use the solution so obtained in estimating the various physical features mentioned above. Other formal methods of analyzing these types of problem have been recently discussed in [3]. Finally, we draw some conclusions about the nanodevice characteristics and possibility of the control of decoherence and correlation in realistic situations. An example of this is an atom in a Paul trap coupled to engineered controllable reservoirs [4].

In the literature, we often find (A) theoretical proposals for future experiments [5, 6] and (B) preliminary experiments [7, 8, 9] on simple coherent systems such as quantum point contacts, quantum dots, and Josephson junctions to examine issues of decoherence and their control. It should be pointed out that in contrast to [4], these do not involve control via the reservoir. In this paper we discuss some of these in the same exploratory spirit.

For simplicity of presentation, we focus our attention on a single, one-dimensional system. We define the time-dependent density matrix in the usual way:

$$\langle x_1 | \rho(t) | x_2 \rangle = \langle x_2 | \rho(t) | x_1 \rangle^* \quad \text{(Hermiticity)}$$

$$Tr\rho(t) = \int dx \langle x|\rho(t)|x\rangle = 1 \quad \text{(Trace class)} \tag{1}$$

$$\iint dx_1 dx_2 \phi^*(x_1)\langle x_1|\rho(t)|x_2\rangle\phi(x_2) \geq 0 \quad \text{(positive definiteness)}$$

Here * stands for complex conjugate. We define the center of mass and relative coordinates, $r_1 + r_2 = 2R$, $r_1 - r_2 = r$, and define the density matrix in the form

$$\langle R + \frac{1}{2}r|\rho(t)|R - \frac{1}{2}r\rangle \equiv \rho(R,r;t) \tag{2}$$

Throughout we use units where the Planck constant, $\hbar = 1$. The "ambiguity function" $A(Q,r;t)$, is defined as the Fourier transform of the density matrix with respect to R, and the "Wigner function" $f(R,p;t)$ as the Fourier transform with respect to r, as follows:

$$\rho(R,r;t) = \int_{-\infty}^{\infty} \frac{dQ}{2\pi} e^{-iQR} A(Q,r;t)$$

$$= \int_{-\infty}^{\infty} \frac{dp}{2\pi} e^{-ipr} f(R,p;t) \tag{3}$$

The properties listed in eq. (1) are reflected as the corresponding properties of the two functions defined above as follows:

$$A^*(Q,r;t) = A(-Q,-r;t), \text{ And } f^*(R,p;t) = f(R,p;t). \tag{3a}$$

And the normalization condition

$$A(Q=0, r=0;t) = 1 = \iint \frac{dR\, dp}{2\pi} f(R,p;t). \tag{3b}$$

Also $\quad f(R,p;t) = \int_{-\infty}^{\infty} dr \int_{-\infty}^{\infty} \frac{dQ}{2\pi} e^{-iQR} e^{ipr} A(Q,r;t). \tag{3c}$

The most general Gaussian form for the density matrix is set up by defining $A(Q,r;t)$ with time-dependent coefficients:

$$A(Q,r;t) = \exp-\left(\frac{1}{2}A(t)r^2 + B(t)rQ + \frac{1}{2}C(t)Q^2 + D(t)r + E(t)Q\right). \tag{4}$$

Here we suppress the time dependence but when we consider the solution of the Lindblad equation, we exhibit this explicitly. The condition (3a) imposes the following requirements on the coefficients in eq. (4) while condition (3b) is fulfilled by construction:

    A, B, C are real, and D, E are pure imaginary.     (5)

An immediate consequence of eq. (4) and the relation to the Wigner function given in eq. (3c), one may derive the following mean values of R, p, and their dispersions and the correlation between R and p, represented by the mean value of the product (Rp) as follows:

$$m(R) = \int_{-\infty}^{\infty}\int_{-\infty}^{\infty} \frac{dRdp}{2\pi} R f(R,p;t) = -i\frac{\partial}{\partial Q} A(Q,r;t)\bigg|_0 = iE;$$

$$m(p) = \int_{-\infty}^{\infty}\int_{-\infty}^{\infty} \frac{dRdp}{2\pi} p f(R,p;t) = i\frac{\partial}{\partial r} A(Q,r;t)\bigg|_0 = -iD;$$

$$\sigma_R^2 = \int_{-\infty}^{\infty}\int_{-\infty}^{\infty} \frac{dRdp}{2\pi} R^2 f(R,p;t) = -\frac{\partial^2}{\partial Q^2} A(Q,r;t)\bigg|_0 = C - E^2 > 0;$$

$$\sigma_p^2 = \int_{-\infty}^{\infty}\int_{-\infty}^{\infty} \frac{dRdp}{2\pi} p^2 f(R,p;t) = -\frac{\partial^2}{\partial r^2} A(Q,r;t)\bigg|_0 = A - D^2 > 0;$$

$$\sigma_{Rp} = \int_{-\infty}^{\infty}\int_{-\infty}^{\infty} \frac{dRdp}{2\pi} Rp f(R,p;t) = \frac{\partial^2}{\partial Q\partial r} A(Q,r;t)\bigg|_0 = -B + DE.$$

(6)

Here all the derivatives of the function $A(Q,r;t)$ are evaluated at Q=0=r, as shown. From these the dispersions about the mean are found to be

$$\Delta_R^2 \equiv \sigma_R^2 - m^2(R) = C > 0;\ \Delta_p^2 \equiv \sigma_p^2 - m^2(p) = A > 0;$$
$$\Delta_{Rp} \equiv \sigma_{Rp} - m(R)m(p) = -B.$$

(6')

It is henceforth convenient to measure all distances from m(R): $\bar{r}_i = r_i - m(R),\ i = 1, 2.$ Then we deduce the density matrix and the Wigner function from eq. (4):

$$\rho(R,r;t) = \frac{1}{\sqrt{2\pi C}} \exp-\left(\frac{\bar{R}^2 + 2iB\bar{R}\bar{r} + \Omega^2 \bar{r}^2 + 2iCm(p)\bar{r}}{2C}\right),$$

(7)

where $\Omega^2 = AC - B^2 \geq 1/4$.

And $$f(R,p;t) = \frac{1}{\sqrt{\Omega^2}} \exp-\left(\frac{A\bar{R}^2 + 2B\bar{R}\bar{p} + C\bar{p}^2}{2\Omega^2}\right)$$ (8)

where $\bar{p} = p - m(p)$.

From eq. (7), it is clear that the properties of the density matrix given in eq. (1) are satisfied if A, C and $(AC - B^2)$ are positive. The positivity of A and C is obvious as they are dispersions of momentum and position whereas that of $(AC - B^2) \geq 1/4$ follows from the Heisenberg inequality. These properties must be maintained when we determine them as solutions of equations arising from the Lindblad equation. From eq. (7), we deduce the length scale, $d(corr)$ over which the spatial correlations decay, defined by

$$\left\langle R+\frac{1}{2}r\left|\rho(t)\right|R-\frac{1}{2}r\right\rangle\bigg|_{r=0} = \left\langle r_1\left|\rho(t)\right|r_1\right\rangle$$
$$= \frac{1}{\sqrt{2\pi C}}\exp-\left(\frac{\bar{r}_1^2}{2C}\right) \Rightarrow d(corr) = \sqrt{2C}. \tag{9}$$

Similarly, we obtain the length scale, $d(decoh)$ over which the system is localized, defined by (apart from an irrelevant phase factor for this consideration)

$$\left\langle R+\frac{1}{2}r\left|\rho(t)\right|R-\frac{1}{2}r\right\rangle\bigg|_{R=0} = \left\langle r_1\left|\rho(t)\right|-r_1\right\rangle$$
$$\propto \exp-\left(\frac{2\Omega^2\bar{r}_1^2}{C}\right) \Rightarrow d(decoh) = \sqrt{\frac{C}{2\Omega^2}}. \tag{10}$$

From eqs. (9,10), we also have the result that the two length scales are related to the dispersion in the center of mass coordinate:

$$d(corr) = \sqrt{2\Delta_R^2}; \quad d(decoh) = \sqrt{\frac{\Delta_R^2}{2\Omega^2}}. \tag{10'}$$

There is another physical picture that emerges by rewriting the expression for the density matrix, eq. (7), in terms of the original coordinate:

$$\left\langle r_1\left|\rho\right|r_2\right\rangle = \frac{e^{-im(p)(\bar{r}_1-\bar{r}_2)}}{\sqrt{2\pi C}}\exp-\frac{1}{2C}\left\{\begin{array}{l}\bar{r}_1^2\left(\frac{1}{4}+\Omega^2+iB\right)+\bar{r}_2^2\left(\frac{1}{4}+\Omega^2-iB\right)\\+\bar{r}_1\bar{r}_2\left(\frac{1}{2}-2\Omega^2\right)\end{array}\right\}. \tag{11}$$

This shows that the density matrix is not proportional to a product of a Gaussian wave function and its complex conjugate showing that there is a correlation between the two coordinates. This feature in the one-particle density matrix is the signature of a "mixed" state of the system as is evident by comparing the density matrix of a harmonic oscillator with frequency $\omega_A$, evaluated at finite temperature, $\beta^{-1}$:

$$\left\langle r_1\left|\rho\right|r_2\right\rangle_{Eqm} = \sqrt{\frac{m\omega_A\tanh(\beta\hbar\omega_A/2)}{2\pi\hbar}}\,e^{-im(p)(\bar{r}_1-\bar{r}_2)}$$
$$\exp-\frac{m\omega_A}{2\hbar}\left\{(\bar{r}_1^2+\bar{r}_2^2)(\coth\beta\hbar\omega_A)-2\bar{r}_1\bar{r}_2(\cosech\beta\hbar\omega_A)\right\} \tag{12}$$

A third length scale, $d^2(mix)$, suggests itself in this version which is identified as

$$d^{-2}(mix) = \frac{(4\Omega^2-1)}{8\Delta_R^2}. \tag{13}$$

These three length scales obey an interesting "additive" composition

$$d^{-2}(decoh) = d^{-2}(corr) + 4d^{-2}(mix). \quad (14)$$

The Lindblad theory gives a formally exact quantum dynamical equation for the time-dependent density matrix, possessing desirable properties of preserving positivity of the underlying density matrix and including the possibility of passage from pure state to mixed state and vice versa. This theory has recently been applied to study practical applications in condensed matter physics and chemistry. A general framework for dealing with dissipative harmonic oscillator in this theory was given by Isar et. al. [3] from which they were able to derive several existing models of dissipation as special cases. In view of its importance and due to lack of methods to solve the Lindblad equation, an action principle was constructed recently [10] as a possible avenue for obtaining approximate solutions. For a discussion of some aspects of decoherence and dissipation using the Lindblad formalism, we may refer to several articles in the book by Giulini et.al. [11], in particular the articles by Joos therein. The solution of the Lindblad equation for the density matrix of a dissipative oscillator has been studied before using the Gaussian ansatz for the Wigner function and the density matrix [3, 10]. These give rise to complicated coupled equations for the coefficients appearing in the Gaussian ansatz. An application to decoherence times and validity of a "Lumped Circuit" description of quantum nanodevices was given recently [12]. In this section, we present the solution in terms of the Gaussian ansatz for the ambiguity function, which yields linear equations for the coefficients in eq. (4), which are then solved in a straightforward way.

The Lindblad equation for a dissipative harmonic oscillator of mass m=1 and frequency $\omega_A$, is written in the form

$$\left\{\partial_t + i\left[-\frac{1}{2}\left(\partial_{r_1}^2 - \partial_{r_2}^2\right) + \frac{\omega_A^2}{2}\left(r_1^2 - r_2^2\right)\right]\right\}\rho(r_1, r_2; t)$$

$$= \begin{bmatrix} -\frac{1}{2}h_{11}(r_1 - r_2)^2 + \frac{1}{2}h_{33}\left(\partial_{r_1} + \partial_{r_2}\right)^2 + h_{13}^{(i)}\left(1 + r_1\partial_{r_2} + r_2\partial_{r_1}\right) \\ + ih_{13}^{(r)}(r_1 - r_2)\left(\partial_{r_1} + \partial_{r_2}\right) \end{bmatrix} \rho(r_1, r_2; t). \quad (15)$$

Here $\partial_t = \partial/\partial t$, $\partial_n = \partial/\partial r_1$, and the h's are magnitudes of the complex coupling strengths appearing in the Lindblad equation and these constants together represent the dissipative processes. For simplicity of presentation, we consider these as time-independent. Using the center of mass and relative coordinates introduced earlier, we deduce the equation for the ambiguity function defined in eq.(3):

$$\left\{\partial_t + \left[-Q\partial_r + \omega_A^2 r\partial_Q\right]\right\}A(Q, r; t) \quad (16)$$

$$= \left[-\frac{1}{2}h_{11}r^2 - \frac{1}{2}h_{33}Q^2 + h_{13}^{(r)}Qr - h_{13}^{(i)}\left(Q\partial_Q + r\partial_r\right)\right]A(Q, r; t).$$

The first term in the right hand side of eq. (16) represents "diffusion" or equivalently "decoherence" and the last term represents "friction". Using the Gaussian form for the ambiguity function, eq. (4), we obtain the following equations for the coefficients appearing therein:

$$\partial_t \begin{pmatrix} D(t) \\ E(t) \end{pmatrix} = \begin{pmatrix} -\Gamma_A & -\omega_A^2 \\ 1 & -\Gamma_A \end{pmatrix} \begin{pmatrix} D(t) \\ E(t) \end{pmatrix}, \qquad (17)$$

where $\Gamma_A = h_{13}^{(i)}$.

These are the equations for the mean relative momentum and the center of mass coordinate of the oscillator and the "friction" constant makes its appearance in these equations. The solutions are linear combinations of $e^{\pm i\omega_A t - \Gamma_A t}$, which are just damped harmonic oscillations in both of these quantities. Furthermore, the dispersions obey the following equations:

$$\partial_t \begin{pmatrix} A(t) \\ B(t) \\ C(t) \end{pmatrix} = \begin{pmatrix} -2\Gamma_A & -2\omega_A^2 & 0 \\ 1 & -2\Gamma_A & -2\omega_A^2 \\ 0 & 2 & -2\Gamma_A \end{pmatrix} \begin{pmatrix} A(t) \\ B(t) \\ C(t) \end{pmatrix} + \begin{pmatrix} h_{11} \\ -h_{13}^{(r)} \\ h_{33} \end{pmatrix} \qquad (18)$$

The solutions, A (t), B (t), and C (t) of this equation can be analytically found by Laplace transform technique and are linear combinations of $e^{\pm i 2\omega_A t - 2\Gamma_A t}, e^{-2\Gamma_A t}$ and driven by the Lindblad couplings. These are then used to obtain parameters associated with decoherence, correlation, and mixing, which give us a clue to controlling these important parameters in terms of the Lindblad coupling strengths introduced in eq. (15). The final expressions are numerically evaluated and we will give graphical display of the solution and interpretation of these features for a model system presently. In making this display, we choose position and momentum variables in dimensionless form so that all the Lindblad parameters have dimensions of energy (recall that we use units with the usual Planck constant is chosen to be unity). The time variable is similarly chosen to be dimensionless by introducing an energy variable, $\omega_A$. We have used simple initial conditions in solving the eq. (18) without any correlation between the coordinate and momentum:

$$A(\tau = 0) = \langle p^2 \rangle_0 = b_1, \; B(\tau = 0) = \langle Rp \rangle = 0, \text{ and } C(\tau = 0) = \langle R^2 \rangle_0 = a_1 \qquad (19)$$

From eqs.(18) and (19), we note the time evolution of the coefficients from the given initial state. The time-dependent solution of eq.(18) should be such that A and C should be nonnegative to preserve their basic definitions given in eq.(6'). This places conditions on the choice of values chosen for the Lindblad coefficients. Another important set of relations are obtained by considering the solutions for asymptotically long times, when they reduce to the stationary solutions of eq. (18). These are obtained by setting the left side of eq.(18) to zero. These solutions are expressed in dimensionless form denoted by tilde, as defined above

$$A(\infty) = \frac{1}{\tilde{\Gamma}_A \left( \tilde{\Gamma}_A^2 + 4 \right)} \left\{ \tilde{h}_{11} \left( \tilde{\Gamma}_A^2 + 2 \right) + 2\tilde{h}_{33} + 2\tilde{\Gamma}_A \tilde{h}_{13}^{(r)} \right\} > 0, \qquad (20a)$$

$$B(\infty) = \frac{1}{\left( \tilde{\Gamma}_A^2 + 4 \right)} \left\{ \tilde{h}_{11} - \tilde{h}_{33} - 2\tilde{\Gamma}_A \tilde{h}_{13}^{(r)} \right\} \qquad (20b)$$

and

$$C(\infty) = \frac{1}{\tilde{\Gamma}_A(\tilde{\Gamma}_A^2 + 4)} \left\{ 2\tilde{h}_{11} + \tilde{h}_{33}(\tilde{\Gamma}_A^2 + 2) - 2\tilde{\Gamma}_A \tilde{h}_{13}^{(r)} \right\} > 0. \tag{20c}$$

If the stationary state is taken to be the thermodynamic equilibrium attained by the oscillator at temperature T, then one may treat the temperature as a controlling parameter. In fact, we find in this case, the Lindblad parameters are determined as follows;

$$\tilde{h}_{11}(Th) = \tilde{h}_{33}(Th) = (\tilde{\Gamma}_A/2) \coth(1/2\tilde{T}), \quad \tilde{h}_{13}^{(r)}(Th) = 0. \tag{21}$$

Here $\tilde{T}$ is the dimensionless temperature parameter. From this and eqs.(10', 13) we deduce the following important length scales:

$$d_{Th}(corr) = (\coth(1/2\tilde{T}))^{1/2}, \quad d_{Th}(decoh) = (\tanh(1/2\tilde{T}))^{1/2},$$
$$\text{and} \quad d_{Th}(mix) = (2\sinh(1/\tilde{T}))^{1/2}. \tag{22}$$

We now present the results of the computation of the solutions graphically and discuss its implications to possible experimental situations mentioned in the Introductory paragraph. In Fig.(1), the dispersions of position, momentum, and their cross-correlation scaled appropriately to make them dimensionless are plotted against dimensionless time for a convenient set of Lindblad parameters. They all reach their asymptotic values given by eqs.(20 a, b, c) for times of about $\tau > 10$. In Fig.2, we display the behavior of decoherence lengths given by eq.(10'), as well as the uncertainty product, normalized by their initial time values, as a function of dimensionless time for three representative values of the Lindblad parameter, $h_{13}^{(r)}$, while holding the other three parameters fixed. Interestingly they all reach the asymptotic value for scaled times of the order of 1. (It should be noted that the asymptotic values of the uncertainty product is off-scale in this figure). Note that the tuning of the decoherence length mirrors the change in the uncertainty product as a function of time. Decoherence lengths larger than unity is obtained for some limited short time scales $(0.1 < \tau < 0.3,$ roughly estimated$)$ for the choice (a) implies that the typical length scale of quantum behavior of the system is lost on these time scales but recovers after those times. In experiments using engineered reservoirs, such as the Paul trap experiments [4], where the environment can be controlled, this type of behavior may be useful to determine and avoid such regimes.

Another way of interpreting this is to estimate the thermal decoherence length from eq.(22) and see when it is reached on this plot. This gives an approximate measure of the operating temperature above which the system would reach nonquantum regime. These types of theoretical estimates, it is hoped may be useful in designing experiments such as those mentioned in [4-9]. These suggest some possible ways of controlling features of decoherence in nanosystems.

 ACKNOWLEDGEMENTS
Both the authors are supported in part by the Office Naval Research. We also thank Dr. Peter Reynolds of the Office Naval Research for supporting this research.

## REFERENCES

[1] G. Lindblad, Commun. Math. Phys. **48**, 119 (1976); see also V. Gorini, A. Kassakowski, and E. C. G. Sudarshan, J. Math. Phys. **17**, 821 (1976).
[2] A. Papoulis, J. Opt. Soc. Am**. 64**, 779 (1974).
[3] A. Isar, A. Sandulescu, and W. Scheid, Phys. Rev. **E60**, 6371 (1999).
[4] C. J. Myatt, B. E. King, Q. A. Turchette, C. A. Sackett, D. Klelpinski, W. M. Itano, C. Monroe, and D. J. Wineland, Nature **403**, 269 (2000).
[5] Y. Makhlin, G. Schön, and A. Shnirman, Nature **398**, 305 (1999); J. E. Mooij, T. P. Orlando, L. Levitov, L. Tian, C. H. van der Wal, and S. Lloyd, Science **285**, 1036 (1999).
[6] B. Elattari and S. A. Gurvitz. Phys. Rev. Lett. **84**, 2047 (2000).
[7] N. H. Bonadeo, J. Erland, D. Gammon, D. Park, D. S. Katzer, and D. G. Steel, Science **282**, 1473 (1998).
[8] Y. Nakamura, Y. A. Pashkin, and J. S. Tsai, Nature **398**, 786 (1999).
[9] M. Switkes, C. M. Markus, K. Campman, and A. C. Gossard, Science **283**, 1905 (1999).
[10] A. K. Rajagopal, Phys. Lett. **A228**, 66 (1997).
[11] D. Giulini, E. Joos, C. Kieffer, J. Kupsch, I. -O. Stamatescu, and H. D. Zeh, *Decoherence and the Appearance of a Classical World in Quantum Theory,* Springer-Verlag, New York (1996).
[12] A. K. Rajagopal, Phys. Lett. **A246**, 237 (1998).
8

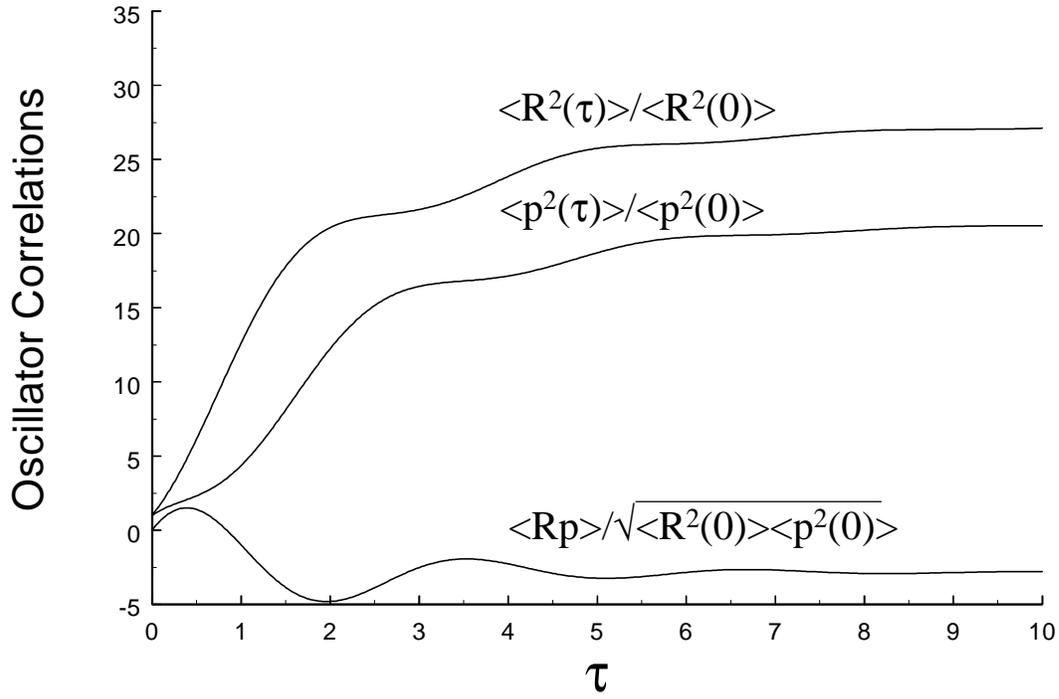

Figure 1. Plot of the analytical solutions of eq.(18) with initial conditions of eq.(19) and $\Gamma_A=0.5$, $h_{11}=4$, $h_{33}=8$, and $h_{13}^{(r)}=4$.

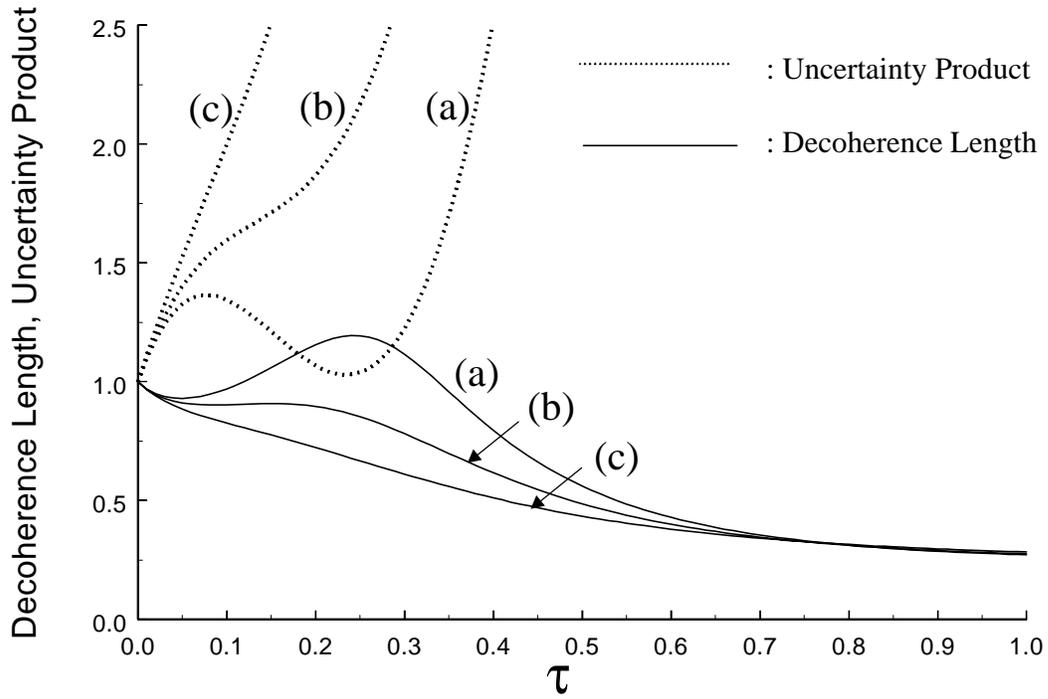

Figure 2. Decoherence length, given by eq.(10'), and the uncertainty product, based on the analytical solutions of eq.(18) with initial conditions of eq.(19) along with minimum uncertainty dispersions. The following values of the parameters were used: $\Gamma_A=0.5$, $h_{11} = 4$, $h_{33} = 8$, and (a) $h_{13}^{(r)} = 4$, (b) $h_{13}^{(r)} = 4.6$, and (c) $h_{13}^{(r)} = 4.9$.